%% file: main.tex
\documentclass{article}

\usepackage[labelfont=bf]{caption}
\usepackage{cite}
\usepackage{hyperref} 
\usepackage{tabularx}
\usepackage{multirow}
\usepackage{adjustbox}
\usepackage{placeins}
\usepackage[a4paper,top=1cm,bottom=2cm,left=3cm,right=3cm,marginparwidth=1.75cm]{geometry}
\usepackage{parskip}
\usepackage{etoolbox}
\usepackage{graphicx}
\usepackage{amsmath}
\usepackage{caption} 
\usepackage{float}
\usepackage{authblk}
\usepackage{xcolor}
 \usepackage{lineno}
\hypersetup{
    colorlinks=true,
    linkcolor=black,
    filecolor=black,      
    urlcolor=black,
    citecolor=black
    }
\patchcmd{\thebibliography}{\chapter*}{\section*}{}{}
\title{\vspace{-1.5cm} Non-invasive optical quantification of methanol in bottled spirits}

\author[1,2]{An\'e Kritzinger}
\author[1]{George O. Dwapanyin}
\author[2]{Ralf Mouthaan}
\author[1,$\dagger$]{Graham D. Bruce}
\author[1,2,$\dagger$,*]{Kishan Dholakia}

\affil[1]{SUPA, School of Physics and Astronomy, University of St Andrews, North Haugh, St Andrews, Fife, UK}
\affil[2]{Centre of Light for Life and School of Biological Sciences, The University of Adelaide, Adelaide, Australia}
\affil[$\dagger$]{These authors contributed equally.}
\affil[*]{kd1@st-andrews.ac.uk}

\date{}

\begin{document}
\maketitle
% \linenumbers
\begin{abstract}
%%% AK
Food and beverage contamination poses a persistent global threat. A prime example is the presence of methanol in counterfeit or illicit spirits, causing severe and often fatal poisoning worldwide. Rapid, non-destructive, and on-site screening methods capable of molecular analysis directly through commercial packaging are therefore urgently needed for quality control and consumer safety. Here, we introduce a non-invasive optical approach based on Raman spectroscopy that judiciously combines wavefront shaping with wavelength modulation to enhance the signal-to-noise ratio and enable quantification of methanol in unopened bottled spirits. A limit of detection of 0.2\% (v/v) methanol in 40\% ethanol was achieved, well below the 2\% (v/v) threshold for safe human consumption. This truly non-invasive method remains robust through coloured glass bottles, with calibration validated in a real spirit sample. By enabling through-container methanol detection, the technique offers a practical tool to protect consumers and streamline routine screening across the beverage supply chain. Moreover, this Raman geometry establishes a versatile platform for assessing authenticity, composition, and contaminants directly through packaging.

\end{abstract}

% %Required by PNAS
% \textbf{Significance statement:} Methanol is a toxic alcohol and is sometimes present in alcoholic drinks, either by accident or deliberately added, posing a serious threat to consumer safety. Existing detection methods typically require opening bottles and laboratory analysis, which is impractical for on-site testing. We developed a non-invasive technique using Raman spectroscopy with wavelength modulation and a specially shaped excitation beam, allowing methanol to be quantified in bottled spirits -- even through commercial coloured glass bottles. This approach achieved a detection limit well below the level dangerous to humans and worked reliably in a real whisky sample. Our method could support faster, safer, and more widespread testing of alcoholic beverages, helping protect consumers and improve quality control in the global beverage industry.


\section*{Introduction}

Food and beverage products are often compromised by contaminants, which can diminish product quality and pose serious risks to human health. A particularly pressing public health threat is the presence of methanol in adulterated or counterfeit alcoholic spirits~\cite{bryan2024worldwide, manning2021illicit, honeyman2025deadly, pressman2020review}. Harmful amounts of methanol can occur in spirits because of poor distillation processes, but more frequently it is illegally added as a cheaper and more readily available substitute for ethanol \cite{nekoukar2021methanol}. The widespread prevalence of counterfeit spirits, estimated to be as high as 40\% of global consumption, underscores the scale of the problem \cite{bryan2024worldwide}. While methanol itself is not toxic, it becomes harmful to humans when metabolised to formaldehyde and formic acid, which can lead to blindness and, in severe cases, to death \cite{jangjou2023awareness}. The medical humanitarian organisation M\'edecins Sans Fronti\`eres tracks suspected methanol poisoning incidents worldwide and reported 31 incidents across 13 countries, resulting in 265 fatalities in just the first half of 2025 \cite{msfOutbreaksWorldwide}. These statistics highlight the need for rapid, cost-effective, and on-line methods for methanol detection.

Chromatography remains the gold standard for methanol detection \cite{botelho2020methanol}. Although highly sensitive, it is destructive, relies on expensive instrumentation operated by trained personnel, and analysis is confined to a laboratory setting. Recent developments in colorimetric and fluorometric sensors based on nanoparticles, organic molecules, and inorganic compounds offer simpler and cheaper alternatives; nonetheless, these methods remain invasive as they require direct contact with the sample \cite{sangeetha2025recent, deoghoria2025precision, behera2024paper}. Handheld gas sensors capable of selectively detecting methanol through headspace (ullage) or breath analysis have also emerged \cite{abegg2020pocket, vandenBroek2019highly, hassan2025machine,tonezzer2022nanosensor}. Despite these advances, there is an absence of non-invasive and non-destructive methods to selectively detect harmful methanol levels in spirits without the need to open the bottle. This gap prevents the implementation of on-line and high-throughput screening at production sites, customs, or retail settings. 

Spectroscopic techniques like infrared (IR) spectroscopy \cite{yang2016determination, power2021what} and Raman spectroscopy (RS) \cite{boyaci2012novel, degoes2016light, ashok2013optofluidic, song2017online, chi2023development} give molecular information and are well established for detecting alcohols in spirits. The non-contact nature of spectroscopy makes it well-suited for through-container analysis. In particular, spatially offset Raman spectroscopy (SORS), a technique in which the excitation and collection regions are physically separated to probe deeper within a sample, has enabled various through-barrier analyses, ranging from detecting explosives in plastic bottles to monitoring glucose levels through the skin \cite{mosca2021spatially, zhang2025subcutaneous, lee2023direct}. A related and potentially more powerful approach uses wavefront shaping to generate a conically shaped excitation beam, which has proven successful for non-invasive authentication of whisky through clear glass bottles \cite{fleming2020through, shillito2022focus}. Unlike other SORS geometries, this method maintains collinearity between the excitation and collection regions while effectively bypassing the container signal \cite{shillito2022focus}. Notably, several RS techniques for ethanol and methanol quantification in spirits through bottles have been proposed \cite{nordon2005comparison, ellis2017through, ellis2019rapid, papaspyridakou2023non, menevseoglu2021non, kiefer2017analysis}. However, all these methods are either limited to clear glass containers or highly dependent on a specific container and spirit matrix, preventing the establishment of a single calibration model for methanol detection and limiting their broader applicability. Although Raman spectroscopy shows the most promise for non-invasive molecular analysis, there remains an opportunity to develop universally applicable methods that are robust across container and sample types, enabling reliable, real-time detection of contaminants such as methanol in sealed products.

As a consequence of optical excitation, fluorescence emission as well as Raman signals are normally recorded. The resultant fluorescence can be orders of magnitude stronger than the underlying Raman signal, thus confounding signal analysis. In our case, fluorescence may originate from both the sample and the container, making this a challenging issue to overcome. Numerous fluorescence removal techniques exist, including post-processing background subtraction methods, photobleaching, selecting an excitation wavelength where fluorescence is minimal, leveraging the different lifetimes of fluorescence emission and Raman scattering, and modulating the excitation wavelength \cite{wei2015review}. Among these, wavelength-modulated Raman spectroscopy (WMRS) is particularly effective and, in contrast to numerical approaches, less prone to artefacts and removal of true Raman features subsumed in a fluorescent background. In WMRS,  the excitation wavelength is tuned, shifting the Raman peaks while the fluorescence background remains constant, allowing their effective separation \cite{de2010online}. WMRS enhances sensitivity by improving the signal-to-noise ratio (SNR) and inherently provides a true baseline correction, avoiding any bias introduced by fitting-based methods. Owing to these advantages, WMRS has been successfully applied across a range of applications, particularly in biomedical research \cite{chen2020optical, baron2020real, woolford2018towards}.

In this paper, we present a novel, non-invasive approach for quantifying methanol in unopened bottled spirits. By combining wavefront shaping and wavelength modulation, our method effectively suppresses the signal from the container while enhancing key Raman features. This calibration strategy remains robust even when applied through coloured glass bottles and in a complex whisky matrix. Our technique reliably detects methanol concentrations well below the maximum tolerable level of 2\% volume by volume (v/v) methanol in 40\% (v/v) ethanol for safe human consumption \cite{paine2001defining}, highlighting its potential for rapid, in-situ quality control in the beverage industry. While we illustrate its use for methanol detection, the approach is broadly applicable and may readily be extended to other contaminants and sample types.

\section*{Results and Discussion}
\subsection*{The non-invasive Raman system}

To measure the Raman spectrum of a sample through a container, three main challenges must be addressed: 1) the spectroscopic signal from the container masking the sample signal; 2) the intrinsic fluorescence signal of the sample that can overwhelm the weaker Raman peaks; and 3) the opacity and colour of the glass attenuating both the excitation beam delivered to the sample and the emitted Raman signal exiting the container. 

\begin{figure}[t!]
    \centering
    \includegraphics[width=\textwidth]{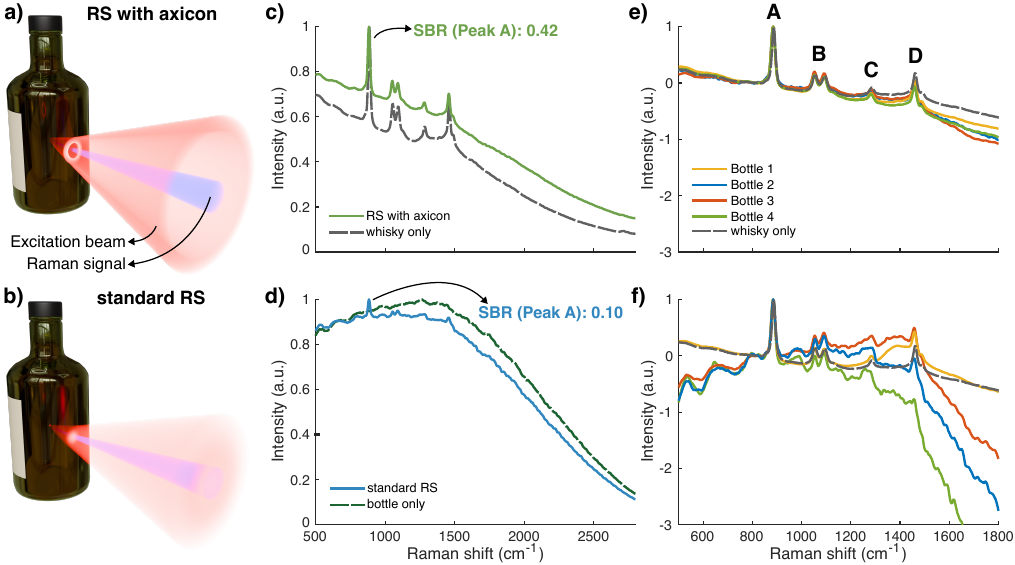}
    \caption{\textbf{Comparing RS with axicon and standard RS.} Conceptual diagrams of the through-bottle Raman geometry using a) a conical excitation beam and b) a Gaussian excitation beam. Example through-bottle spectra of whisky (Whisky~4) acquired with c) the RS with axicon configuration and d) the standard RS configuration. The spectrum collected with the axicon geometry is dominated by the signal of the contents, giving a signal-to-background (SBR) of 0.42 for Peak~A. In contrast, the standard RS spectrum is dominated by the fluorescence of the container, with only small Raman peaks visible (SBR = 0.10). For reference, c) and d) include a spectrum of the whisky measured directly through a vial (no bottle) and the spectrum of the empty bottle (Bottle~4), respectively. Raman spectra of Whisky~4 measured through different bottles (Bottles 1-4) using e) RS with axicon and f) standard RS. The axicon-based measurements are more consistent and primarily reflect the signal from the contents, while the standard RS spectra vary more due to the influence of the different containers. The spectra in e) and f) were normalised to the main ethanol peak (Peak A) to account for intensity variations caused by the different containers.}
    \label{Axicon setup: Gauss vs axicon}
\end{figure}

% \subsubsection*{The cone-shaped excitation beam}
To address the first challenge, a conically shaped excitation beam is generated using an axicon lens \cite{fleming2020through}. This wavefront-shaped beam forms an annular beam on the surface of the bottle and subsequently collapses to a focus within the bottle. The Raman signal of the liquid sample is excited at the focal point inside the bottle and then collected in a back-scattered configuration through the middle `dark' region of the annular beam, thus avoiding the signal from the container, as shown schematically in Fig.~\ref{Axicon setup: Gauss vs axicon}a. This geometry is referred to as \textit{RS with axicon}. See Materials and Methods, and Supplementary Fig.~S1 for a detailed description of the optical setup. 

The advantage of using this conically shaped excitation beam is evident when compared to a standard Gaussian excitation beam (referred to as \textit{standard RS}) -- illustrated in Fig.~\ref{Axicon setup: Gauss vs axicon}b. The spectrum of a commercial whisky was measured through its coloured glass bottle with both the wavefront-shaped and Gaussian excitation beams, as shown in Fig.~\ref{Axicon setup: Gauss vs axicon}c and \ref{Axicon setup: Gauss vs axicon}d. Our axicon system effectively circumvents the bottle signal, enabling measurement of the whisky signal and significantly improving the signal-to-background ratio (SBR). In contrast, the spectrum obtained using the standard beam is dominated by the fluorescence signal originating from the bottle, with only small Raman peaks from the sample superimposed on this background. In Fig.~\ref{Axicon setup: Gauss vs axicon}e and \ref{Axicon setup: Gauss vs axicon}f, the spectrum of the same whisky was measured through four different glass bottles with the two configurations. When using the axicon setup, the signal of the sample was consistently obtained regardless of the bottle it was placed in, contrary to the measurements done with a standard Gaussian beam, where the fluorescence signal of the different bottles contributed more to the measured signal. 

Typically, due to the high percentage of ethanol, the Raman spectra of spirits are mostly dominated by the Raman signal of ethanol, sometimes superimposed on a fluorescence background for spirits with more complex matrices like whisky (see Fig.~\ref{Axicon setup: Gauss vs axicon}e). The main Raman peak of ethanol (Peak A) at 886~cm$^{-1}$ is attributed to C--C stretching, Peak B (1053 and 1097~cm$^{-1}$) is from C--O stretching and C--H rocking, Peak C at 1277 cm$^{-1}$ is due to C--H torsion and rotation and finally Peak D at 1457~cm$^{-1}$ is from C--H bending \cite{salinas2023raman, papaspyridakou2023non, boyaci2012novel}. 

The conical excitation beam, therefore, effectively reduces interference from the container (Challenge~1). Considering the second and third challenge -- autofluorescence of the sample and attenuated signals due to the container -- the most straightforward strategy is careful selection of the excitation wavelength. Ideally, the excitation beam is readily transmitted through the bottle while minimising fluorescence emission from the sample. In practice, however, transmission varies greatly across containers of different colours and materials, making it difficult to identify a universally optimal wavelength (see Supplementary Fig.~S2). To reduce fluorescence emission, longer wavelengths in the near-infrared region are typically used \cite{wei2015review}. Accordingly, a 785 nm laser was used in this study, which proved to have sufficient transmission through coloured glass bottles while limiting fluorescence from complex whisky matrices (see the Supplementary Figs.~S2, S3 and S4 for a detailed discussion on the choice of excitation wavelength). Nonetheless, wavelength selection alone cannot fully overcome sample fluorescence while maintaining optimal transmission through the bottles. To further suppress fluorescence and enhance sensitivity, we combined the wavefront-shaped excitation beam with wavelength-modulated Raman spectroscopy~(WMRS).

% \subsubsection*{Wavelength modulation}

\begin{figure}[t!]
    \centering
    \includegraphics[width=\textwidth]{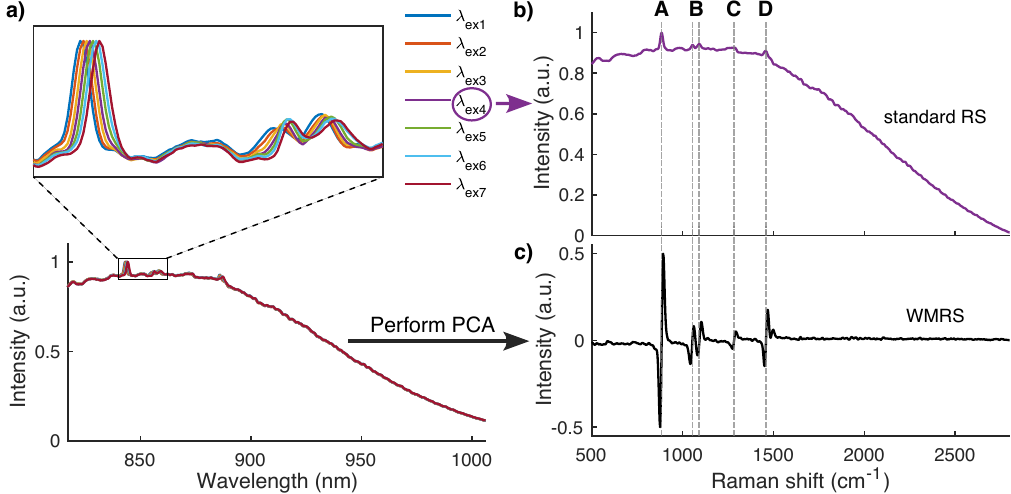}
    \caption{\textbf{WMRS principle and analysis.} In WMRS, multiple Raman spectra are recorded while changing the wavelength of the excitation light ($\lambda_{\text{ex}}$). a) Seven Raman spectra of whisky (Whisky 4) were measured through its bottle as $\lambda_{\text{ex}}$ is varied over a range of 1~nm. For small wavelength changes, the Raman peaks shift accordingly while the background fluorescence stays constant. In b), a single Raman spectrum excited at the central wavelength ($\lambda_{\text{ex4}}$) is shown. c) Principal component analysis (PCA) performed on the seven spectra identifies a first principal component (PC1) with loadings which indicate the position and relative intensity of the Raman peaks. The Raman peak positions correspond to the zero-crossing of the peaks in the WMRS spectrum, and the relative Raman peak intensities are retained.  Importantly, this new spectrum does not contain features from wavelength-independent fluorescence, meaning Raman peaks that are difficult to identify with standard RS (consider Peak~C) become more evident with WMRS. } 
    \label{WMRS setup: explain}
\end{figure}

WMRS exploits the fact that Raman peaks shift relative to the excitation wavelength, whereas the fluorescence emission remains largely unchanged for small (a few nanometres) excitation wavelength shifts. By modulating the wavelength during measurement, the Raman peak positions can be extracted while suppressing the background fluorescence signal. The principle and analysis of WMRS is illustrated in Fig.~\ref{WMRS setup: explain}.  

To obtain the wavelength-modulated spectrum, the excitation wavelength was tuned in seven equidistant steps over a range of 1~nm. The resulting spectra of whisky measured through its bottle at each excitation wavelength are shown in Fig.~\ref{WMRS setup: explain}a, where the Raman peaks shift with wavelength while the fluorescence background stays constant. Several analysis methods can be used to extract the Raman peak positions from these spectra, including standard deviation analysis, least-squares fitting, and principal component analysis (PCA) \cite{mazilu2010optimal}. PCA is a widely used unsupervised statistical method that extracts the maximum variation between the spectra, the Raman peaks in our case, which is given by the first principal component (PC1).  A typical WMRS spectrum obtained using PCA is presented in Fig.~\ref{WMRS setup: explain}c, where the zero-crossings indicate the Raman peak positions, matching those in the standard Raman spectrum acquired at the central excitation wavelength (Fig.~\ref{WMRS setup: explain}b). Notably, this WMRS spectrum contains only Raman features, with the fluorescence background directly removed, avoiding any bias and spectral artefacts that numerical fitting might introduce. PCA was selected over the other analysis methods because it provides the highest signal-to-noise ratio (SNR) while preserving the relative Raman peak intensities \cite{mazilu2010optimal}.  The enhanced sensitivity of WMRS over standard Raman is evident from Peak C, which is barely discernible in the standard spectrum but appears clearly in the WMRS spectrum.

The effect of the wavelength step size and total modulation range on the SNR of the resulting WMRS spectrum is shown in Supplementary Fig.~S5. Increasing the number of excitation wavelengths and broadening the modulation range both improve the SNR, but at the cost of longer acquisition times. Due to the low transmission of some containers tested in this study, an integration time of 5~s with 5 accumulations per excitation wavelength was necessary, yielding a total acquisition time of 25~s for each standard spectrum. A step size of 0.17~nm over a modulation range of 1~nm was chosen, producing seven standard spectra for each WMRS measurement. This resulted in a total acquisition time of approximately 180~s per WMRS spectrum, which was sufficient to achieve robust detection while maintaining experimental feasibility.

In the simplest implementation of WMRS, where only two excitation wavelengths are used, the method is equivalent to shifted-excitation Raman difference spectroscopy (SERDS) \cite{gebrekidan2016shifted}. While using more excitation wavelengths generally improves the SNR of the resulting WMRS spectrum (see Supplementary Fig.~S5), SERDS can still provide effective fluorescence suppression with an appropriately chosen wavelength shift, as illustrated in Supplementary Fig.~S6. The optimal step size for SERDS depends on factors such as the natural spacing of the Raman peaks of interest, the spectrometer resolution, and the wavelength and linewidth of the excitation laser. Although our method was not optimised for speed in this study, it may readily be made faster ($\sim$10~s) for on-line analysis by increasing the laser power and reducing the integration time, number of accumulations, or the number of excitation wavelengths. These parameters can be specifically optimised for methanol detection, which would make the method faster, more compact and practical for high-throughput screening (see Supplementary Fig.~S7 for a compact device design).

% \subsubsection*{System performance}

\begin{figure}[b!]
    \centering
    \includegraphics[width=\textwidth]{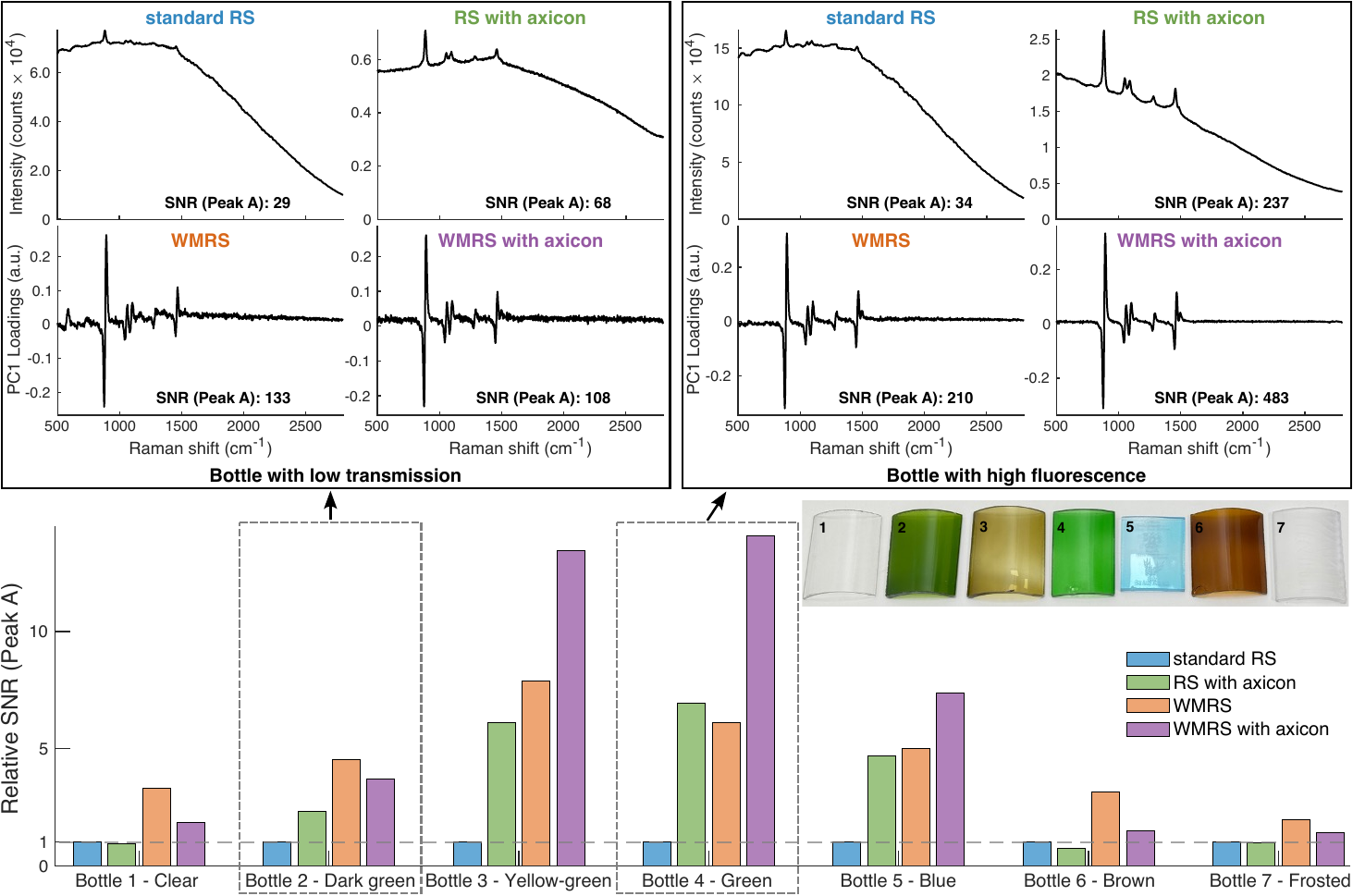}
    \caption{\textbf{Improved SNR in through-bottle Raman spectroscopy using wavefront shaping and wavelength modulation.} The four methods (standard RS, RS with axicon, WMRS, and WMRS with axicon) were used to acquire the Raman spectra of commercial spirits through their sealed bottles. The relative SNRs obtained with each method for all the containers are represented in the bar graph, normalised to the SNR of the standard RS for clarity. Representative spectra obtained using the four configurations through two of the containers (Bottles 2 and 4) are displayed above. The inset is a photo of cut-outs (65~mm $\times$ 55~mm) of the bottles used in this experiment. }
    \label{SNR: compare all methods}
\end{figure}

In this study, we combined WMRS with the conically shaped excitation beam. The system can therefore operate in four modes: 1) \textit{standard RS} using a Gaussian excitation beam at a single wavelength, 2) \textit{RS with axicon} using a conical excitation beam at a single wavelength, 3) \textit{WMRS} using a Gaussian excitation beam at multiple wavelengths, and 4) \textit{WMRS with axicon} which uses the conical excitation beam at multiple wavelengths. To evaluate the performance of each method, Raman spectra of sealed spirits were measured directly through their bottle, and the SNR of the main ethanol peak (Peak A) was compared (Fig.~\ref{SNR: compare all methods}). As shown in this figure, seven different bottles of varying colour and translucency were investigated. Details of the spirits and the spectroscopic properties of the glass bottles are presented in Supplementary Tab.~S1 and Fig.~S8.  In most cases, using the conical excitation beam improved the SNR relative to standard RS. WMRS enhanced the SNR for all the containers, and the combination of WMRS and wavefront shaping improved the SNR significantly for three of the samples. 

The variation in the performance of the conical excitation beam between containers can be attributed to the optical properties of the glass. This conical beam mitigates fluorescence emission from the glass, thereby improving SNR in bottles where the limiting factor is the high fluorescence emission of the bottles (Bottles~3 -- 5). For bottles with a lower fluorescence emission, but limited transmission of the excitation light (Bottles~2 and 6), standard WMRS outperformed the WMRS with axicon system. At the same power, the standard Gaussian excitation beam produces a higher signal than the conical excitation beam, even though the SBR is lower (Supplementary Fig.~S9). As low transmission limits the detected signal, a higher raw signal leads to a higher SNR. The reason for the decreased signal for the axicon configuration is the presence of the iris in the collection path, which partially blocks the signal. Additionally, the focal spot formed inside the bottle was more distorted for the conically shaped beam than for the standard Gaussian beam (Supplementary Fig.~S10), further reducing signal intensity. Overall, through-bottle Raman measurements consistently benefit from wavelength modulation, and the added use of the wavefront-shaped excitation beam is particularly advantageous for highly fluorescent containers.  

These results demonstrate that the judicious combination of wavelength-modulation with the conical excitation beam can greatly improve the sensitivity of obtaining the Raman signal through commercial coloured glass bottles, necessary for detecting low methanol concentrations.

\subsection*{Methanol detection}

\begin{figure}[hb!]
    \centering
    \includegraphics[width=\textwidth]{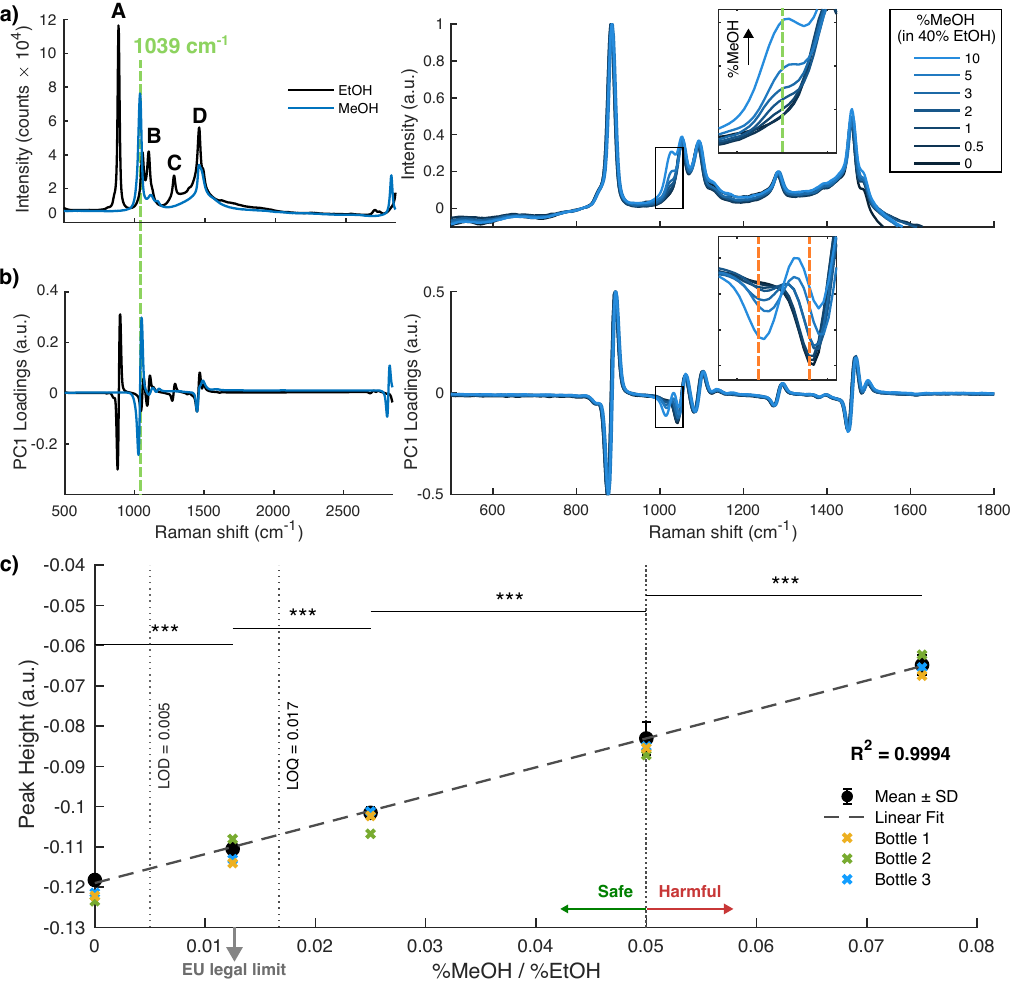}
    \caption{\textbf{Calibration graph for methanol detection.} a) Standard Raman spectra and b) corresponding WMRS spectra of pure ethanol (EtOH) and methanol (MeOH) (left), and mixtures of 40\% (v/v) ethanol with increasing methanol concentration (right). Zoomed-in insets highlight the spectral region used to extract the methanol peak height. At low concentrations, negative peak heights arise in the WMRS spectra from calculating the peak-to-peak distance (orange lines). c) Calibration regression curve showing a linear relationship between methanol peak height and methanol concentration, measured using the WMRS with axicon configuration. Each point is the average of five replicate measurements through Bottle~4; measurements through three additional bottles (Bottles~1 -- 3) are also shown. The LOD and LOQ are indicated on the graph, along with the legal EU limit and the maximum tolerable threshold for safe human consumption. Statistical significance between neighbouring calibration points was assessed with an unpaired t-test (* P$<$0.05, ** P$<$0.01, and *** P$<$0.001).}
    \label{MeOH: Calibration graph}
\end{figure}

The main advantage of using Raman spectroscopy for methanol detection in spirits is its ability to selectively detect methanol in the presence of ethanol, as some of their characteristic Raman peaks are spectrally distinct (Fig.~\ref{MeOH: Calibration graph}a). The most intense methanol peak at 1039~cm$^{-1}$ is attributed to C--O stretching, and appears as a shoulder on ethanol Peak B.

The transmission of laser light through different commercial spirit bottles varies widely, depending on the glass colour, thickness (which can vary around the bottle circumference), and the bottle shape. This makes the actual power interrogating the sample unknown, which complicates absolute quantification of analytes through sealed bottles. Instead, relative peak intensities in the Raman spectra can be used to determine analyte concentrations. In this study, we assume the ethanol concentration of the sample is known and use it as an internal standard to quantify the methanol. Notably, ethanol content in counterfeit products is often overestimated because dilution is a common form of fraud \cite{bryan2024worldwide, menevseoglu2021non}. Consequently, any inaccuracy in the assumed ethanol concentration would more likely lead to an overestimation of methanol levels, which is preferable to underestimation from a consumer safety perspective.

To establish a calibration curve for quantifying methanol through bottles, solutions containing 40\% (v/v) ethanol with varying methanol concentrations were prepared. Raman spectra of these solutions were measured through Bottle~4 with all four configurations. Bottle~4 was selected as the most optically challenging container due to its high fluorescence emission and low transmission. Fig.~\ref{MeOH: Calibration graph}a~and~b present representative through-bottle standard and WMRS spectra of these solutions. The methanol concentration was obtained by extracting the peak height of the main methanol Raman band (position indicated in the insets) relative to the main ethanol peak (Peak A). 

The resulting calibration curve obtained with the WMRS with axicon geometry is shown in Fig.~\ref{MeOH: Calibration graph}c. The limit of detection (LOD) and limit of quantification (LOQ) were calculated as $3\cdot\sigma_{\text{blank}}/m$ and $10\cdot\sigma_{\text{blank}}/m$, respectively, where $\sigma_{\text{blank}}$ is the standard deviation of the blank measurements ($\text{n}=5$) and $m$ is the slope of the calibration graph. Calibration parameters for the other methods are summarised in Table~\ref{tab: calibration} (see Supplementary Fig.~S11 for the calibration curves). Standard RS did not exhibit a linear response between the methanol peak height and concentration due to strong interference from the container. In contrast, the other three methods demonstrated a good linear response, with the combination of WMRS and the conically shaped excitation beam achieving the lowest LOD. Additional measurements taken through three other bottles (Bottles~1--3) are also shown on the calibration plot and align with the points obtained through Bottle~4, demonstrating the robustness of the calibration across different containers.

\begin{table}[ht!]
\centering
\caption{Calibration performance for all four Raman configurations, showing the coefficient of determination (R$^2$, linearity) and limit of detection. }
\label{tab: calibration}
\begin{tabular}{|c|c|c|}
\hline
\textbf{Method} & \textbf{R$^2$} & \textbf{LOD} \\ \hline
standard RS & 0.6106 & NA \\
RS with axicon & 0.9932 & 0.013 \\
WMRS & 0.9895 & 0.012 \\
WMRS with axicon & 0.9994 & 0.005 \\ \hline
\end{tabular}
\end{table}

Notably, the LOD and LOQ values are below the maximum methanol concentration for safe human consumption (2\% (v/v) methanol in 40\% ethanol \cite{paine2001defining}). The EU legal limit, set at 10~g of methanol per litre of pure ethanol for spirits (Regulation EU 2019/787), corresponds to a methanol-ethanol ratio of 0.014 in 40\% ethanol \cite{okolo2023recent, botelho2020methanol}. This limit lies between the LOD and LOQ, demonstrating that the method reliably quantifies methanol at concentrations relevant to consumer safety, while still offering indicative measurements close to the legal limit.

The method was validated in a real sample matrix by spiking a commercial whisky (Whisky~4) with known concentrations of methanol, see Table~\ref{tab: recovery}. For concentrations above the LOQ, the method achieved satisfactory recoveries ranging from 104\% to 108\%. This validation assumed that the ethanol concentration reported on the bottle was correct, as it is a reputable brand, and that the whisky contained no methanol; however, methanol may be naturally present, which could account for the slight overestimation observed. By demonstrating robust calibration and accurate quantification through bottles with challenging optical properties and in a real whisky matrix, this novel method proves to be independent of both the container and the sample composition.

\begin{table}[htb]
\centering
\caption{Recovery of methanol in spiked whisky (40\% v/v ethanol). Spiked concentration, measured concentration, and percentage recovery are reported, demonstrating accurate and reliable quantification of methanol through a coloured glass bottle.}
\label{tab: recovery}
\begin{tabular}{|c|c|c|}
\hline
\multirow{2}{*}{Spiked \% methanol} & Measured \% methanol & \multirow{2}{*}{Recovery (\%)} \\
 & (mean $\pm$ SD, n=5) &  \\ \hline
0.52 & 0.68 $\pm$ 0.2 & 131 \\
1.0 & 1.1 $\pm$ 0.1 & 108 \\
2.0 & 2.1 $\pm$ 0.2 & 106 \\
3.0 & 3.2 $\pm$ 0.1 & 104 \\ \hline
\end{tabular}
\end{table}

\FloatBarrier
\section*{Conclusion}

In conclusion, a novel non-invasive method based on Raman spectroscopy was introduced for quantifying methanol through unopened spirit bottles using ethanol as an internal standard. Wavefront-shaping of the excitation beam ensured suppression of the fluorescence from the glass container, while wavelength modulation improved the sensitivity of the system, enabling robust calibration and reliable analysis through coloured bottles. A LOD of 0.2\% (v/v) and a LOQ of 0.68\% (v/v) in 40\% ethanol were achieved. This approach could help reduce the risk of methanol poisoning by enabling routine, on-site screening of sealed spirits, and it could be further developed into a compact, field-deployable device. Beyond this specific application, our study uncovered insights into how container properties influence optical measurements, guiding the design of future through-packaging spectroscopic analyses. More broadly, the combined approach of wavefront shaping and WMRS establishes a versatile, non-invasive platform for probing chemical composition in packaged products across food, beverage, pharmaceutical, and cosmetic industries, to support authenticity verification, quality control, and consumer safety.

\FloatBarrier
\section*{Materials and Methods}

\subsection*{Samples and standard solutions}

For this study, seven commercially available spirits were selected, each bottled in different glass containers with challenging optical properties. The details of the spirits are given in Supplementary Tab.~S1, and the transmission and fluorescence emission of the glass bottles are presented in Supplementary Fig.~S8. 

To construct the calibration graph (Fig. \ref{MeOH: Calibration graph}), samples were prepared using standard solutions of 0, 0.5, 1, 2, and 3\% (v/v) methanol ($\geq$99\% purity, Fisher Scientific, UK) and 40\% (v/v) ethanol (99.96\% purity, VWR International, France) with deionised water from a Millipore Elix 5 system. The method was also validated using realistic samples by spiking Whisky~4 with known amounts of methanol. 

\subsection*{Optical setup}

The axicon-based Raman system described by Fleming \textit{et al.} \cite{fleming2020through} was adapted to include wavelength-modulation capability. A schematic of the full setup is shown in Supplementary Fig.~S1. In short, a wavelength-tunable Ti:Sapphire laser source (Coherent Verdi 532nm pumping a Spectra-Physics 3900s), centred at 785~nm, was directed through an axicon lens (AX2510-B, $10^{\circ}$  angle, Thorlabs) to create a Bessel beam. This beam was relayed to the sample plane by lenses L2 (100~mm) and L1 (40~mm), resulting in an annular beam incident on the surface of the bottle before it focuses to a point inside the bottle (15~mm from the outer wall of the bottle). The Raman signal of the contents inside the bottle was collected in a backscattered configuration through the centre of the cone-shaped excitation beam and directed to a spectrometer (Andor Shamrock SR-303i-B with DU920P-BR-DD Newton CCD Camera) via a dichroic mirror (Semrock LPD02-785RU). An iris (with a 12~mm diameter) was added in the collection path to eliminate the signal of the bottle that is excited by the incident annular beam. The effect of the iris and the depth of the focus inside the bottle is shown in Supplementary Fig.~S12. An excitation wavelength of 785~nm was selected for this study, primarily to minimise fluorescence emission from the whiskies while ensuring adequate transmission through green glass bottles (see Supplementary Figs.~S2, S3, and S4).

By replacing the axicon lens with a conventional lens of appropriate focal length (35~mm), a standard Gaussian beam is focused into the bottle rather than the conical beam. The iris is removed for standard Gaussian beam measurements, as the excitation and collection regions overlap completely, making elimination unnecessary. The optical setup can therefore operate under four configurations: standard RS, RS with axicon, WMRS, and WMRS with axicon.

\subsection*{Raman measurements}

The through-bottle Raman spectra were obtained by averaging five spectra, each recorded with an integration time of 5~s and an input slit width of 200~$\mu$m. All spectra were acquired using the Andor Solis software. Replicate measurements were taken at different positions around the bottle, while avoiding the label and glass seams. The excitation laser power was kept at 450~mW at the sample plane, resulting in 90 -- 400~mW focused inside the liquid sample, depending on the specific bottle. The laser power was decreased for bottles with high fluorescence intensity to avoid saturating the detector. For WMRS measurements, the excitation wavelength was stepped seven times over a range of 1~nm, meaning each WMRS spectrum consists of 7 standard Raman spectra.

\subsection*{Spectral processing}

All processing and analysis were performed using MATLAB (R2024b). For WMRS, seven Raman spectra were measured while varying the excitation wavelength over a range of 1~nm. Each spectrum was normalised to the maximum to account for any laser power fluctuations during the modulation. Matlab's standard PCA algorithm was used to obtain the loadings of the first principal component, which retains the positions (at the zero crossings) and relative intensities of the Raman peaks while suppressing the background fluorescence \cite{mazilu2010optimal}. An example of the SNR (peak height to the standard deviation of the noise) and SBR (peak height to fluorescence background) calculations for a standard Raman spectrum and a WMRS spectrum is shown in Supplementary Fig.~S13. The standard deviation of the noise is calculated using the silent Raman region where no Raman peaks are present (1750 -- 2200~cm$^{-1}$) after subtracting a fitted smoothed curve to flatten the spectra, if needed. For the standard spectra, the fluorescence background serves as a baseline from which the Raman peak heights are measured. For the WMRS spectra, the peak height is the distance from the local minimum to the maximum \cite{praveen2012fluorescence}. The SBR is calculated for standard spectra as a measure of the Raman signal to the background fluorescence level. Both SNRs and SBRs are calculated for a specific Raman peak.

\section*{Acknowledgements}
The authors acknowledge Alexander Trowbridge for creating the 3D concept figures (Figs. 1a, 1b, and S7). This work was performed in part at the University of Adelaide OptoFab hub of the Australian National Fabrication Facility, utilising Commonwealth and SA State Government funding, for preparing the glass sections. The authors acknowledge funding support from the University of St Andrews Impact and Innovation Fund,  EPSRC Impact Accelerator - University of St Andrews -  EP/X525819/1, the ARC Laureate Fellowship (grant FL210100099), and the University of Adelaide Research Scholarship (119858)

\section*{Author contributions}
KD and GDB conceived and supervised the study. AK performed the experimental measurements with input from GOD. Data analysis was conducted by AK with input from RPM, GDB, and KD. AK prepared the original draft of the manuscript, and all authors contributed to reviewing and editing the manuscript. All authors approved the final version.

% \FloatBarrier
\bibliographystyle{unsrt}
\small
\bibliography{ReferencesAK.bib}

\include{SI}

\end{document}

%% file: SI.tex
% \documentclass{article}

% % Useful Packages
% % { Packages
% \usepackage[labelfont=bf]{caption}
% \usepackage{cite}
% \usepackage{hyperref} 
% \usepackage{tabularx}
% \usepackage{multirow}
% \usepackage{adjustbox}
% \usepackage{placeins}
% \usepackage[a4paper,top=1cm,bottom=2cm,left=3cm,right=3cm,marginparwidth=1.75cm]{geometry}
% \usepackage{parskip}
% \geometry{margin = 25 mm}
% \linespread{1.5}
% \usepackage{etoolbox}
% \usepackage{graphicx}
% \setlength\parindent{0pt}
% \usepackage{amsmath}
% \usepackage{caption} 
% \usepackage{float}
% \usepackage{authblk}
% % \captionsetup[table]{skip=10pt}
% \hypersetup{
%     colorlinks=true,
%     linkcolor=black,
%     filecolor=black,      
%     urlcolor=blue,
%     citecolor=black
%     }
% \patchcmd{\thebibliography}{\chapter*}{\section*}{}{}
% %}

% \renewcommand{\thefigure}{S\arabic{figure}}
% \renewcommand{\theequation}{S\arabic{equation}}
% \renewcommand{\thetable}{S\arabic{table}}

% \title{\vspace{-1.5cm} Non-invasive optical quantification of methanol in bottled spirits}

% \author[1,2]{An\'e Kritzinger}
% \author[1]{George O. Dwapanyin}
% \author[2]{Ralf Mouthaan}
% \author[1,$\dagger$]{Graham D. Bruce}
% \author[1,2,$\dagger$,*]{Kishan Dholakia}

% \affil[1]{SUPA, School of Physics and Astronomy, University of St Andrews, North Haugh, St Andrews, Fife, UK}
% \affil[2]{Centre of Light for Life and School of Biological Sciences, The University of Adelaide, Adelaide, Australia}
% \affil[$\dagger$]{These authors contributed equally.}
% \affil[*]{kd1@st-andrews.ac.uk}

% \date{}

% \begin{document}
% \maketitle

\renewcommand{\thefigure}{S\arabic{figure}}
\renewcommand{\theequation}{S\arabic{equation}}
\renewcommand{\thetable}{S\arabic{table}}
\setcounter{figure}{0}
\setcounter{table}{0}
\setcounter{equation}{0}

\section*{\huge\centering Supplementary Information}

\begin{table}[h]
\caption{Details of the commercially available spirit samples and their containers used in this study.}
\begin{adjustbox}{width=1\textwidth}
\label{tab: samples}
\begin{tabular}{|cc|ccccc|}
\hline
\multicolumn{2}{|c|}{\textbf{Container}} & \multicolumn{5}{c|}{\textbf{Contents}} \\ \hline
\textbf{Name} & \textbf{Colour} & \textbf{Spirit name} & \textbf{Spirit type} & \textbf{Brand} & \textbf{Volume (mL)} & \textbf{\%EtOH} \\ \hline
Bottle 1 & Clear & Whisky 1 & Blended Scotch whisky & Famous Grouse & 700 & 40 \\
Bottle 2 & Dark green & Whisky 2 & Single malt Scotch whisky & Ardbeg (The Ultimate)& 700 & 46 \\
Bottle 3 & Green-yellow & Whisky 3 & Single malt Scotch whisky & Caol Ila (12 years) & 700 & 43 \\
Bottle 4 & Green & Whisky 4 & Single malt Scotch whisky & Glenfiddich (original 12 years) & 700 & 40 \\
Bottle 5 & Blue & Gin 1 & Gin & Bombay Sapphire & 700 & 40 \\
Bottle 6 & Brown & Whisky 5 & Blended Canadian whisky & Canadian Club 1858 & 700 & 40 \\
Bottle 7 & Frosted glass & Vodka 1 & Vodka & 24 seven vodka & 700 & 40 \\ \hline
\end{tabular}
\end{adjustbox}
\end{table}

\begin{figure}[htbp]
    \centering
    \includegraphics[width=0.75\textwidth]{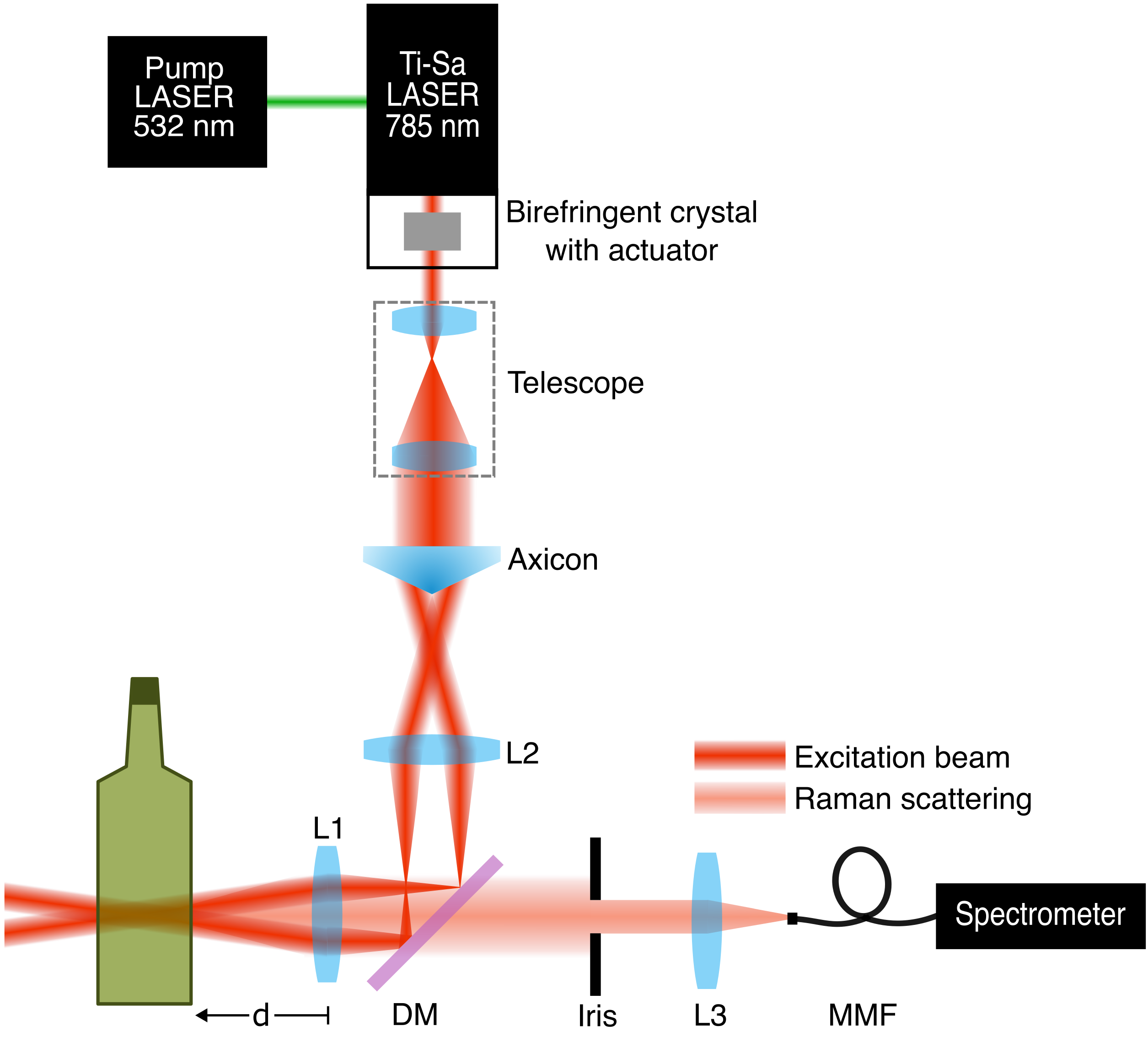}
    \caption{\textbf{A schematic of the system used to measure the Raman spectrum of spirits in a sealed bottle.} A wavelength-tunable laser source, centred at 785~nm, is directed through an axicon lens to create a Bessel beam. This beam is relayed to the sample plane by lenses L2 and L1, resulting in an annular beam incident on the surface of the bottle before it focuses to a point inside the bottle. The Raman signal of the contents inside the bottle is collected in a backscattered configuration through the `dark' centre of the excitation beam and directed to a spectrometer via a dichroic mirror. An iris is added in the collection path to eliminate the signal of the bottle that is excited by the incident annular beam.  MMF: multi-mode fibre; DM: dichroic mirror; L1: lens (40~mm); L2: lens (100~mm) and L3: lens (50~mm).}
    \label{Setup: WMRS with axicon}
\end{figure}

\FloatBarrier
\subsection*{Choice of the excitation wavelength}
\begin{figure}[htbp]
    \centering
    \includegraphics[width=\textwidth]{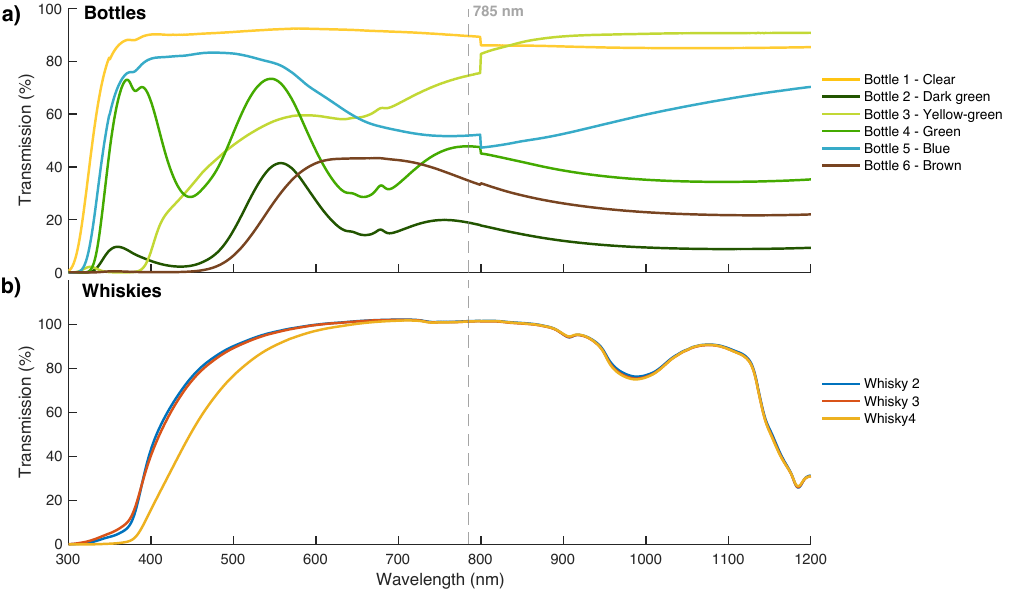}
    \caption{\textbf{Transmission spectra of the containers and contents.} The transmission spectrum of a) six glass containers and b) three whiskies. The excitation wavelength used in this study, 785~nm, is highlighted on the graphs. All transmission spectra were recorded using an Agilent Cary 5000 UV-Vis-NIR Spectrophotometer.}
    \label{Fig: transmission bottles of whiskies}
\end{figure}

Several factors should be considered when choosing the wavelength of the excitation source for non-invasive Raman spectroscopy. Firstly, the excitation beam should be transmitted through the container wall; the transmission spectra of the different glass bottles used in this study are shown in Fig.~\ref{Fig: transmission bottles of whiskies}a. Since these bottles differ significantly in colour, the transmission differs significantly at different wavelengths. The wavelength used in this study, 785~nm, is indicated on the graph. Secondly, the excitation should be transmitted through the liquid deep enough to form the focus inside the sample; this is not a big issue for spirits (see the transmission spectra of whisky in Fig.~\ref{Fig: transmission bottles of whiskies}b), however must be considered, for example, when analysing wine samples.

\begin{figure}[htbp]
    \centering
    \includegraphics[width=\textwidth]{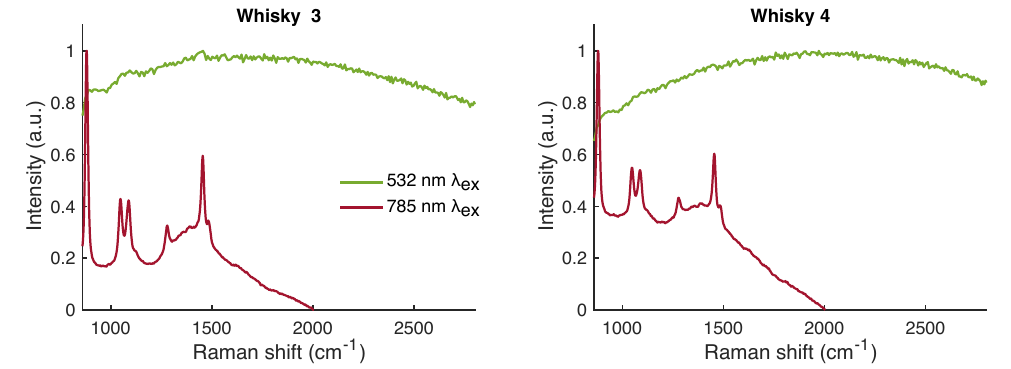}
    \caption{\textbf{Raman spectra of whiskies at different excitation wavelengths.} The Raman spectra of Whisky~3 and 4 excited with a 532~nm and 785~nm laser source. At 532~nm the fluorescence signal from the whisky dominates, whereas the Raman signal dominates at 785~nm. }
    \label{Fig: ex wavelength fluo}
\end{figure}

Since this work focuses on collecting the Raman signal of the sample, it is important to consider the fluorescence emission of the sample at different excitation wavelengths. The emission spectra of two whiskies measured with a 532~nm and 785~nm excitation laser are shown in Fig.~\ref{Fig: ex wavelength fluo}. Both 532~nm and 785~nm are commonly used Raman excitation wavelengths. It is clear that at a lower excitation wavelength, the fluorescence of the whisky dominates and masks the Raman peaks, whereas the Raman peaks are visible with the 785~nm near-IR excitation wavelength. 

\begin{figure}[htbp]
    \centering
    \includegraphics[width=\textwidth]{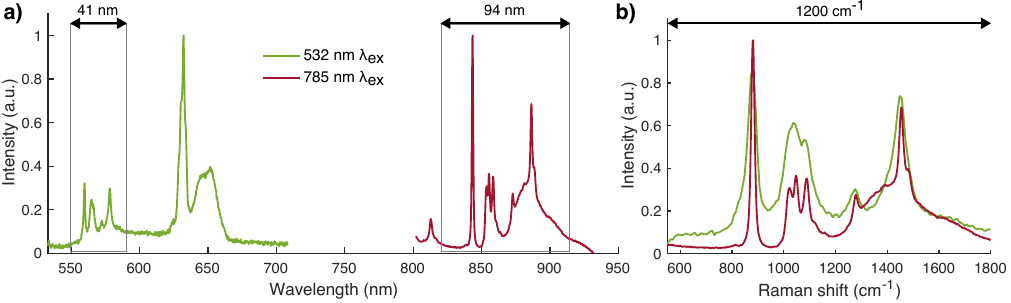}
    \caption{\textbf{Raman spectra at different excitation wavelengths.} The Raman spectra of a 10\% (v/v) methanol and 40\% (v/v) ethanol mixture measured with a 532~nm and 785~nm excitation wavelength plotted in a) wavelengths (nm) and b) Raman shift (cm$^{-1}$). At lower excitation wavelengths, the Raman peaks are more closely spaced, resulting in a lower resolution Raman spectrum and the need for a higher resolution spectrometer.  }
    \label{Fig: ex wavelength res}
\end{figure}

A final aspect to consider is the inherent nature of Raman spectroscopy. Firstly, Raman scattering efficiency scales inversely with the fourth power of the excitation wavelength, meaning shorter wavelengths produce stronger signals, but often at the cost of increased fluorescence (Fig.~\ref{Fig: ex wavelength fluo}). Secondly, Raman peaks are more closely spaced at shorter excitation wavelengths, as shown in Figure \ref{Fig: ex wavelength res}. The methanol peak of interest, which coincides with the doublet peak of ethanol, is therefore more easily separated at longer excitation wavelengths; however, additional factors such as the spectrometer resolution and laser linewidth would also influence peak separation.\\

\begin{figure}[htbp]
    \centering
    \includegraphics[width=\textwidth]{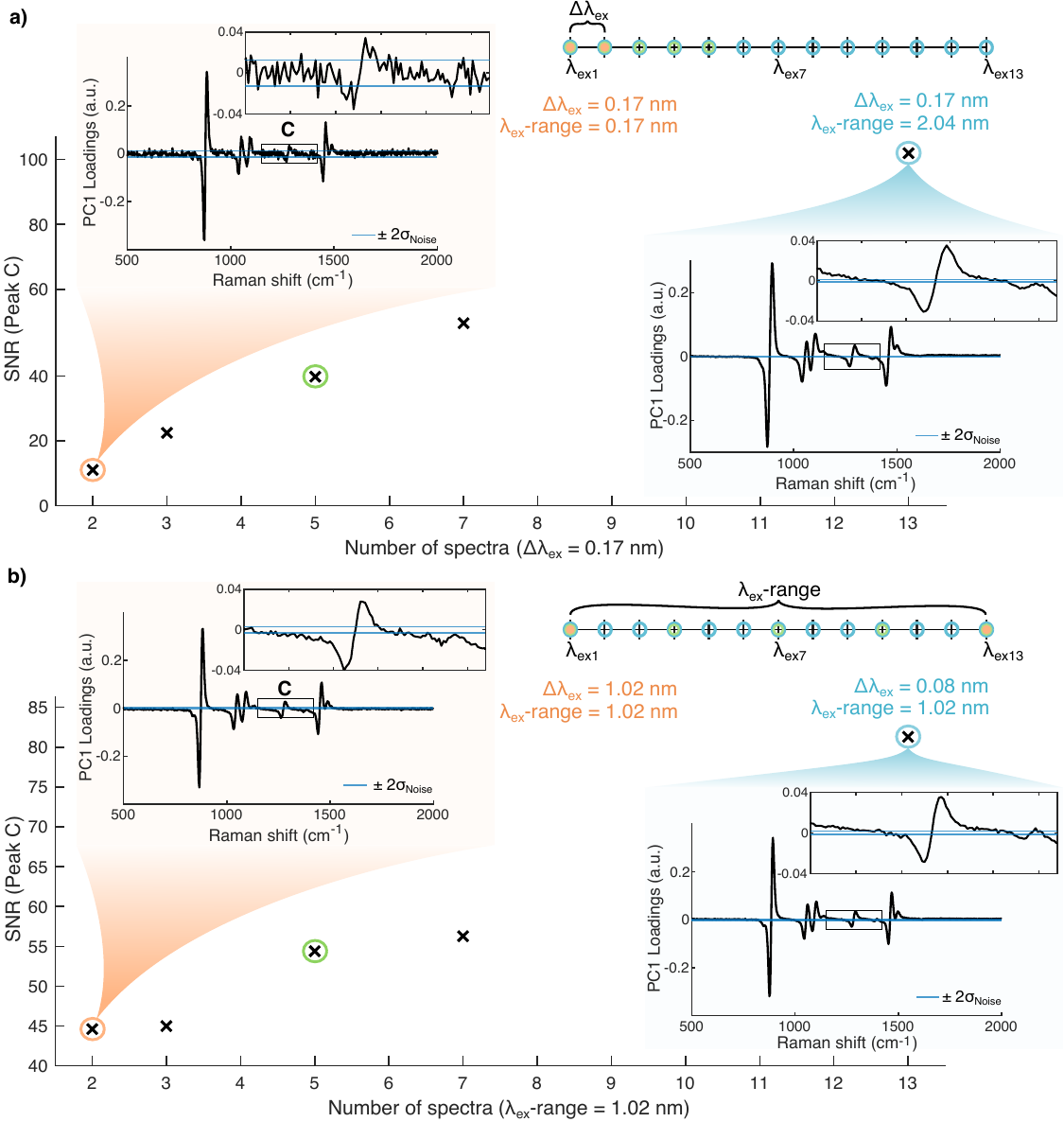}
    \caption{\textbf{Wavelength-modulation parameters.} The WMRS spectrum of Whisky 3 was measured through the bottle while increasing the number of excitation wavelengths used under two conditions: a) a fixed wavelength step size ($\Delta\lambda_{\text{ex}}$) and b) a fixed modulation range ($\Delta_{\text{ex}}$-range). When the step size is constant, the modulation range increases with increasing number of spectra; and when the range is fixed, the step size decreases with increasing number of excitation wavelengths. The insets show the measured WMRS spectra obtained when using 2 and 13 excitation wavelengths. The wavelength distributions for 2, 5, and 13 wavelengths are shown in the bar insets on each graph in orange, green, and blue, respectively. The results indicate that increasing the modulation range has a greater impact on the signal-to-noise ratio (SNR) than increasing the number of acquired spectra within the range. When only two excitation wavelengths are used, the spectrum is equivalent to shifted-excitation Raman difference spectroscopy (SERDS).}
    \label{WMRS: stepsize tests}
\end{figure}

\begin{figure}[htbp]
    \centering
    \includegraphics[width=\textwidth]{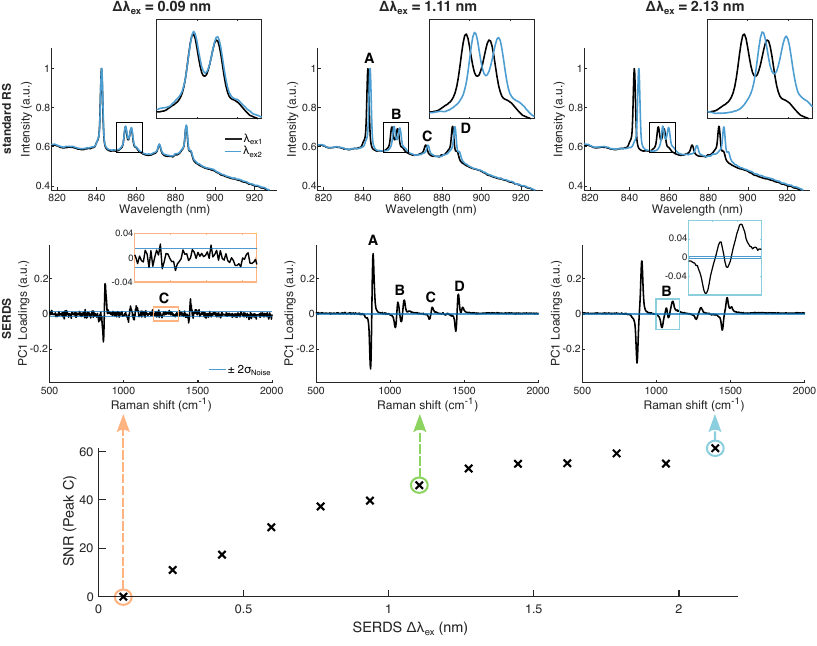}
    \caption{\textbf{Step size for SERDS.} In SERDS, two Raman spectra are acquired at different excitation wavelengths (first row) and then subtracted to obtain the SERDS spectrum (second row). The SNR of Peak C as a function of the wavelength step size is presented in the bottom panel. As the step size increases, the SNR improves but starts to plateau at larger shifts. When the wavelength step is too small ($\Delta\lambda$ = 0.09~nm), the SNR is insufficient to resolve Peak C. In contrast, when the step is too large ($\Delta\lambda$ = 2.13~nm), the resulting SERDS spectrum is distorted, especially in the region of closely spaced Raman peaks, such as the doublet Peak B, which makes interpreting the SERDS spectrum difficult. Selecting an appropriate step size for SERDS can therefore enhance the SNR efficiently without the need to scan across multiple wavelengths, thus simplifying and accelerating measurements compared with WMRS. The sample used in this experiment was Whisky 3.}
    \label{SERDS test}
\end{figure}

\begin{figure}[htbp]
    \centering
    \includegraphics[width=\textwidth]{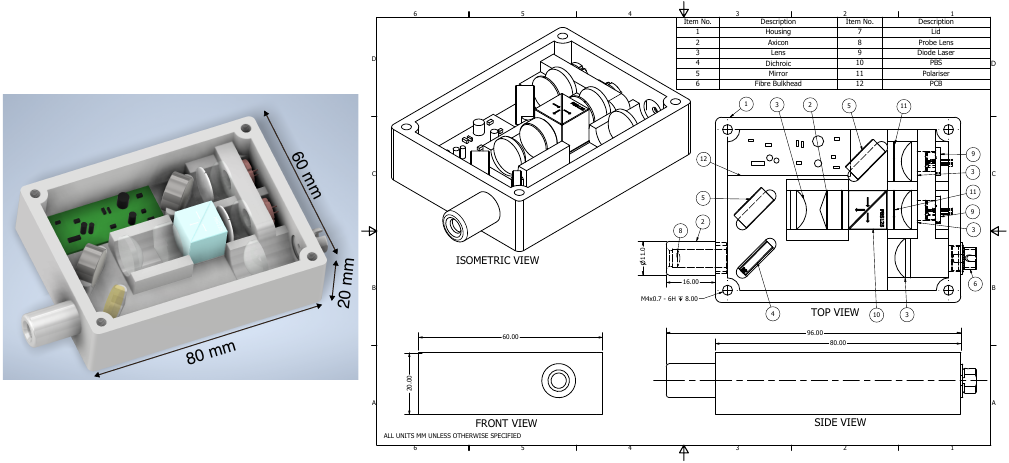}
    \caption{\textbf{Compact design for the through-bottle setup.} A schematic that illustrates a potential arrangement using two diode lasers at different wavelengths for a shifted excitation Raman difference spectroscopy (SERDS) configuration combined with the axicon geometry, highlighting the possibility of a compact implementation. The spectrometer is not included in this design; see the full optical setup in Fig.~\ref{Setup: WMRS with axicon} for an explanation of the role of each component. }
    \label{Compact design}
\end{figure}

\begin{figure}[htbp]
    \centering
    \includegraphics[width=\textwidth]{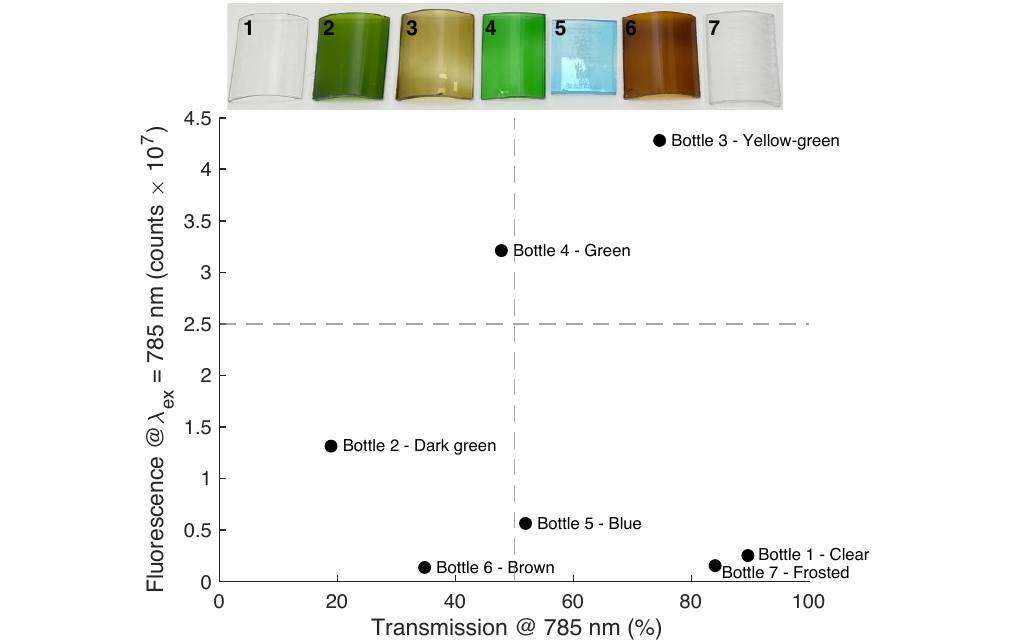}
    \caption{\textbf{Container characterisation.} The integrated fluorescence intensity emitted by each bottle when excited with 785~nm light and the corresponding transmission of the glass at this wavelength. The inset shows a photo of the glass cut-outs (about 65~mm $\times$ 55~mm) of each bottle (Bottle 1--7) used in this study. The transmission was measured with an Agilent Cary 5000 UV-Vis-NIR Spectrophotometer.}
    \label{Fig: Bottle characteristics}
\end{figure}

\begin{figure}[htbp]
    \centering
    \includegraphics[width=0.95\textwidth]{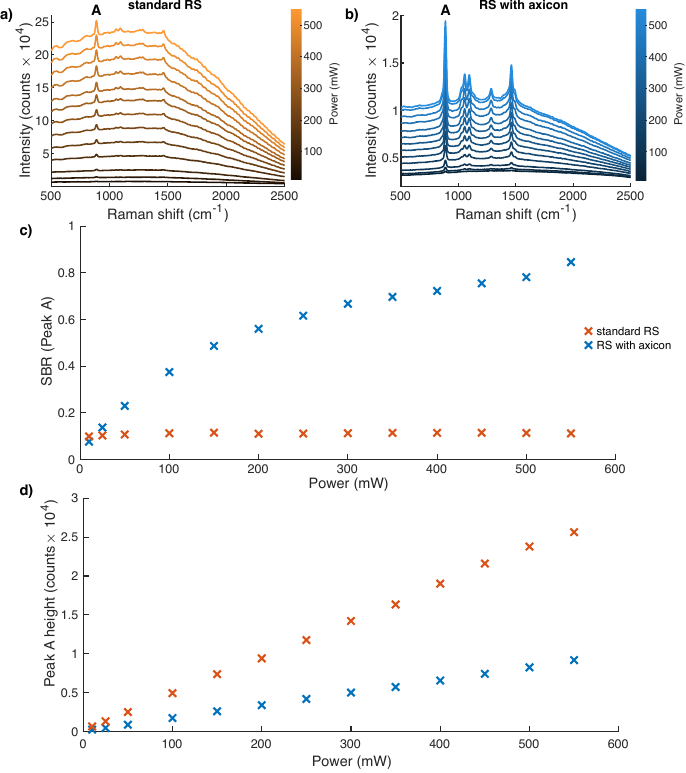}
    \caption{\textbf{Comparing the signal-to-background ratio (SBR) and raw Raman signal of standard RS and RS with axicon.} The Raman spectrum of ethanol through Bottle 4 was acquired using a) standard RS and b) RS with axicon at increasing excitation laser powers. The resulting SBR and peak height of Peak A are plotted for both methods, in c) and d), respectively. In standard RS, the SBR remains nearly constant with increasing power, indicating that both the Raman signal of the contents and the container background scale similarly. In contrast, for the axicon geometry, the Raman signal from the contents increases more significantly than the background at higher powers, resulting in an improved SBR. Even though the axicon system improves the SBR, the raw signal (Peak A height) is much lower than in the standard geometry. This reduced signal arises from partial blocking of the collected Raman light by the iris present in the RS with axicon setup and from a more distorted focus formed inside the bottle for the conical beam compared to the standard beam.}
    \label{Axicon setup: SBR as power change}
\end{figure}

\begin{figure}[htbp]
    \centering
    \includegraphics[width=\textwidth]{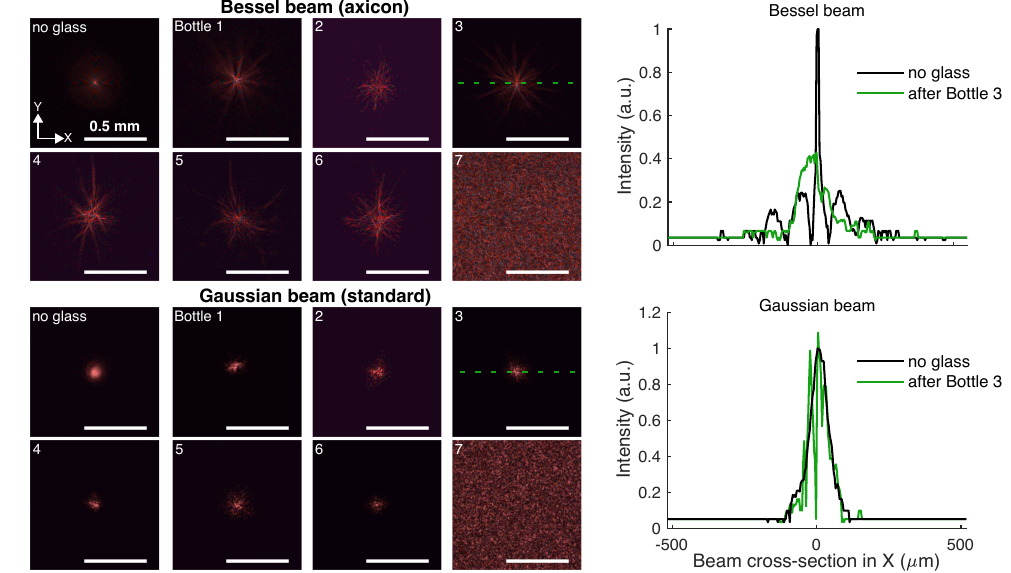}
    \caption{\textbf{Beam profiles of the focal spot inside different bottles.} The focal point formed with the axicon setup (Bessel beam) and with the standard setup (Gaussian beam) was imaged through the sections of glass cut-outs of the different spirit bottles (Bottles 1 to 7). The images were normalised to visualise the different distortions, as the transmission at 785~nm varied between the bottles. On the right, the cross-section of the Bessel and Gaussian beam imaged through Bottle 3 is plotted relative to the beams without the glass piece. The focus created by the axicon system is distorted more by the bottle than the Gaussian beam, lowering the maximum intensity of the focus, which would decrease the Raman signal excited at this point. The frosted glass bottle (Bottle~7) completely distorted the excitation beam, preventing the formation of a focal point within the liquid. Although a signal was obtained through this bottle, better analysis through such highly scattering containers would likely require more advanced wavefront shaping techniques to refocus the beam behind the frosted glass. }
    \label{Fig: beam profiles after glass}
\end{figure}

\begin{figure}[htbp]
    \centering
    \includegraphics[width=0.85\textwidth]{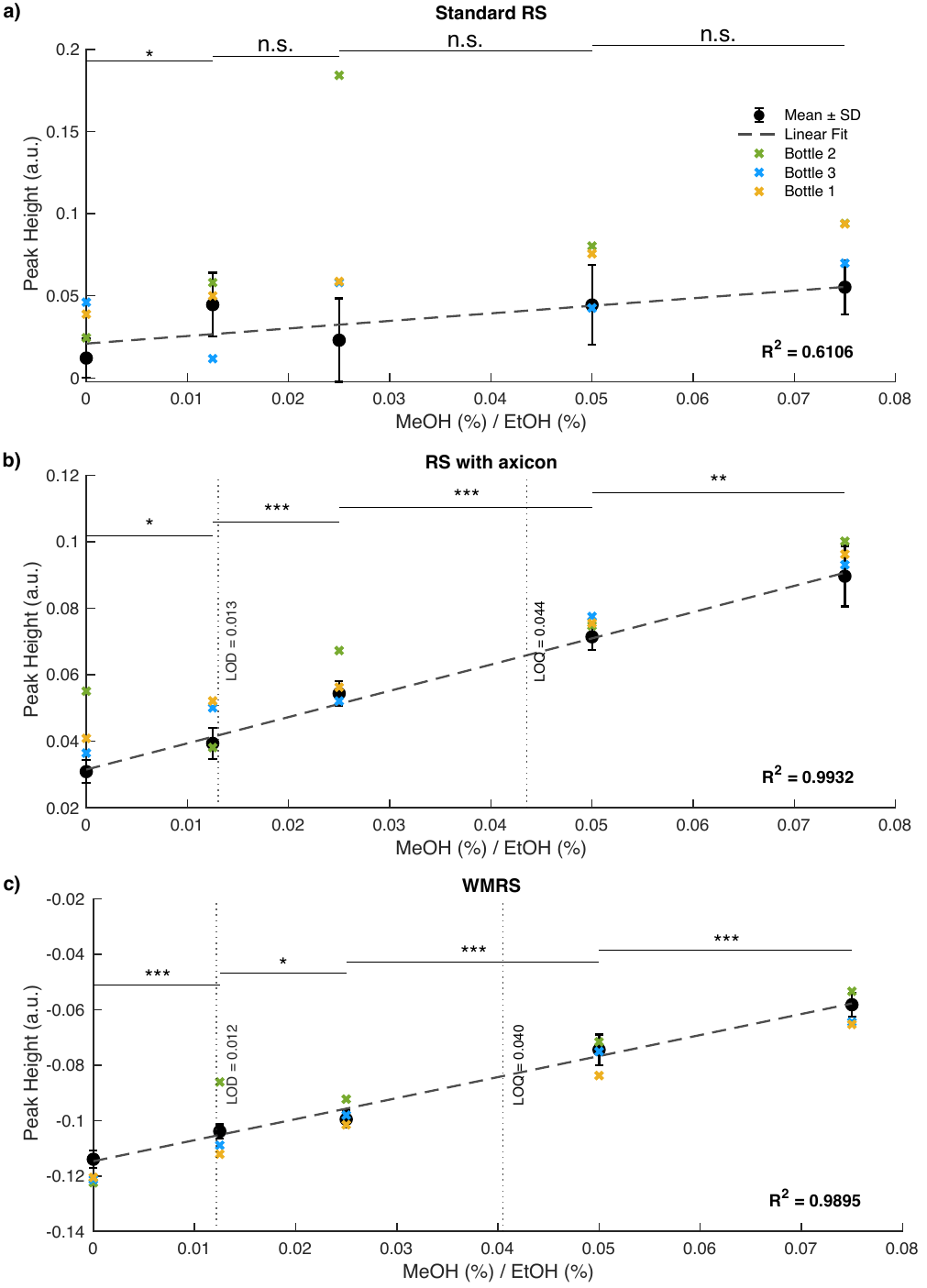}
    \caption{\textbf{Calibration graphs for methanol detection with different configurations.} Calibration regression curve using a) standard RS, b) RS with axicon, and c) WMRS. Each point is the average of five replicate measurements through Bottle~4; measurements through three additional bottles (Bottles~1 -- 3) are also shown. The LOD and LOQ are indicated on b) and c). Since Standard RS did not exhibit a sufficient linear response between the methanol peak height and concentration, the LOD was not calculated. Statistical significance between neighbouring calibration points was assessed with an unpaired t-test (* P$<$0.05, ** P$<$0.01, *** P$<$0.001, and not significant (n.s.) P$\geq$0.05). The calibration graph obtained with the WMRS with axicon geometry is presented in the main paper. }
    \label{Fig: other cal graphs}
\end{figure}

\begin{figure}[htbp]
    \centering
    \includegraphics[width=\textwidth]{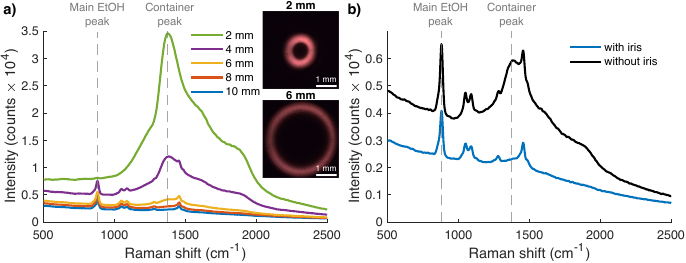}
    \caption{\textbf{Effect of the focus depth and iris in the axicon setup.} a) The Raman spectra of Whisky 1 acquired at various focal depths of the excitation beam inside the bottle (achieved by moving the bottle relative to lens L1). When the focus is positioned 2~mm from the surface, it remains within the bottle wall, resulting in a spectrum dominated by the container signal. As the focus moves deeper into the sample, the contribution from the bottle is progressively suppressed, and the signal from the contents becomes dominant. The insets show the beam profile on the surface of the bottle at different focus depths. b) An iris is required in the collection path to eliminate the container signal excited by the annular beam incident on the bottle. Bottle 1 (clear glass) was used as it provides a narrowly defined container peak, making parameter effects easier to demonstrate.} 
    \label{Axicon setup: distance iris}
\end{figure}

\FloatBarrier

\begin{figure}[htbp]
    \centering
    \includegraphics[width=\textwidth]{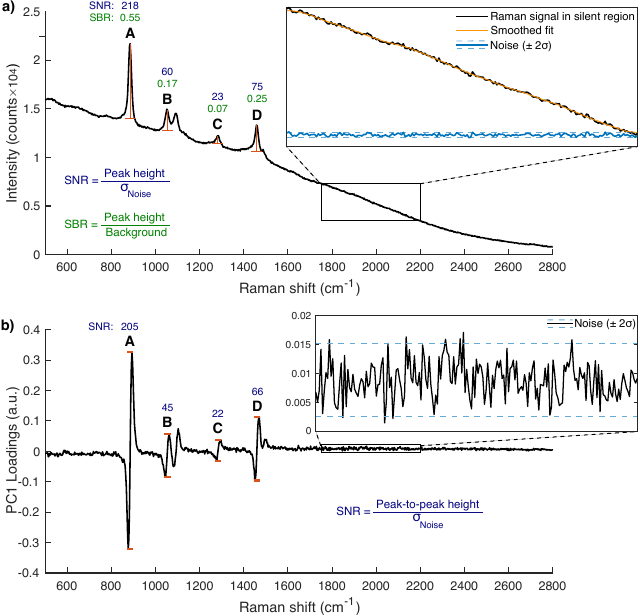}
    \caption{\textbf{Signal-to-noise ratio (SNR) and signal-to-background ratio (SBR) calculations.} An example of the SNR calculation is shown for a) a standard Raman spectrum and b) a WMRS spectrum. The standard deviation of the noise ($\sigma_{\text{Noise}}$) is calculated using the silent Raman region where no Raman peaks are present (1750 -- 2200~cm$^{-1}$) after subtracting a fitted smooth curve to flatten the spectra, if needed. For the standard spectra, the fluorescence background serves as a baseline from which the Raman peak heights are measured. For the WMRS spectra, the peak height is the distance from the local minimum to the maximum of each peak. The SBR is calculated for standard spectra as a measure of the Raman signal to the background fluorescence level. Both SNR and SBR can be calculated for individual Raman peaks; in this study, the ethanol Raman peaks A - D are used for analysis.}
    \label{SNR: calculation}
\end{figure}

% \FloatBarrier
% \bibliographystyle{unsrt}
% \small
% \bibliography{ReferencesAK.bib}

% \end{document}